\pdfoutput=1
\documentclass[referee]{raa}            % referee version: for submission
\usepackage{graphicx,times}   %for PS/EPS graphics inclusion, new
\usepackage{natbib}
\usepackage{enumerate}

\usepackage{amssymb,amsmath}

\bibpunct{(}{)}{;}{a}{}{,}

\usepackage[pagebackref=true]{hyperref}

\begin{document}

  \title{Two-dimensional modeling of the tearing-mode-governed magnetic reconnection in the large-scale current sheet above the two-ribbon flare
}
  %\subtitle{Turbulence in the flare current sheet}

   \volnopage{Vol.0 (20xx) No.0, 000--000}      %%preserved for Editor. DOn't remove!
   \setcounter{page}{1}          %%starting page, preserved for Editor. DOn't remove!

   \author{Yining Zhang  % Put your Chinese name in "( )" if you like. Note to open line 11 "\usepackage[UTF8]{ctex}"
      \inst{1,2}
   \and Jing Ye
      \inst{1,3}
   \and Zhixing Mei
      \inst{1,3}
\and Yan Li
\inst{1,3}
\and Jun Lin
\inst{1,2,3}
   }

   \institute{Yunnan Observatories, Chinese Academy of Sciences, Kunming, Yunnan 650216, People's Republic of China; {\it yj@ynao.ac.cn}\\
        \and
             University of Chinese Academy of Sciences, Beijing 100049, People's Republic of China\\
        \and
             Center for Astronomical Mega-Science, Chinese Academy of Sciences, Beijing 100012, People's Republic of China\\
\vs\no
   {\small Received 20xx month day; accepted 20xx month day}}

\abstract{ We attempt to model magnetic reconnection during the two-ribbon flare in the gravitationally stratified solar atmosphere with the Lundquist number of $S=10^6$ using 2D simulations. We found that the tearing mode instability leads to the inhomogeneous turbulence inside the reconnecting current sheet (CS) and invokes the fast phase of reconnection. Fast reconnection brings an extra dissipation of magnetic field which enhances the reconnection rate in an apparent way. The energy spectrum in the CS shows the power-law pattern and the dynamics of plasmoids governs the associated spectral index. We noticed that the energy dissipation occurs at a scale $l_{ko}$ of 100-200~km, and the associated CS thickness ranges from 1500 to 2500~km, which follows the Taylor scale $l_T=l_{ko} S^{1/6}$. The termination shock(TS) appears in the turbulent region above flare loops, which is an important contributor to heating flare loops. Substantial magnetic energy is converted into both kinetic and thermal energies via TS, and the cumulative heating rate is greater than the rate of the kinetic energy transfer. In addition, the turbulence is somehow amplified by TS, of which the amplitude is related to the local geometry of the TS. 
}

   \authorrunning{Y. Zhang et al.}            %author_head in even pages
   \titlerunning{Turbulence in the flare current sheet }  % title_head in odd pages

   \maketitle
%% The author head (on even pages) and the title head (on odd pages) will be
%% automatically extracted from \author{} and \title{}. Whenever the title is too long,
%% you will be asked to supply a shorter one by inserting either \authorrunning{} or
%% \titlerunning{} before \maketitle. Anyway, you can specify your own heads.
%%
%%
%% Note: In the following text body of your manuscript, please note several differences from
%%       other major journals:
%% (1) \subsection{Please Capitalize the First Letter of Each Notional Word in Subsection Title}
%% (2) Please Capitalize the First Letter of Each Notional Word in all tables' captions

%
%________________________________________________ sections below
%
\section{Introduction}           %% first-level sections will be auto-capitalized
\label{sec:intro}

Solar flares are the most violent events in the solar system which are involved in the conversion of the magnetic energy up to $10^{27}-10^{32}$ ergs. Magnetic reconnection plays a key role in this process and in helping convert the magnetic energy into heating and kinetic energy of plasma, and in accelerating charged particles. Magnetic reconnection process also widely exists in astrophysical studies including solar atmosphere, earth magnetosphere \citep{Priest2000}, black hole accretion disk \citep{Yuan2009,Yuan2012,Meng2015} and magnetic neutron stars \citep{Meng2014}. 

Several types of magnetic reconnection (MR) exist in the solar activities. \citet{Parker1957} and \citet{Sweet1958} described a very long and thin diffusion region of MR, which can only be used to explain slow energy releasing events. \citet{Petschek1964} introduced a single X-type reconnection site combined with slow mode shocks in the outflow regions, in order to explain the fast MR process. Recently, turbulence has gained much attention on what kind of role it plays in MR process. \citet{Lin2007} and \citet{Loureiro2007} pointed out that the turbulence in MR can effectively accelerate energy dissipation in the thick CS. Traditional theories \citep{Petschek1964} imply that the energy is transferred from large scales to small scales and finally dissipated at the ion inertial scale, which is tens of metres in the coronal environment. However, \citet{Forbes1991} and \citet{Riley2007} pointed out that the tearing mode instability plays a 
key role in magnetic diffusion and governs the CS thickness. \citet{Lazarian2020} suggested that turbulence requires the energy to cascade into smaller scales. The fragmented CSs and plasmoids in 2D can be classified into turbulence, while the inverse cascade of merging loops is not. How turbulence thickens CS and accelerates reconnection is quantified by \citet{Lazarian1999} with theoretical predictions supported
by numerical simulations \citep{Kowal2009}. This means that the real thickness of the CS and diffusion scale could be much larger than the ion inertial scale. 

The work by \citet{Lin2007} shows the thickness of CS up to $6.4\times 10^4$~km. \citet{Ciaravella2008} deduced the thickness from the UVCS data in high temperature spectral lines [Fe $\uppercase\expandafter{\romannumeral18}$] and [Ca $\uppercase\expandafter{\romannumeral14}$] and the value is $2.8\times 10^5$~km. Many observations support that the CS width reaches a quite large scale in contrary with the classical theories \citep{Savage2010,Lin2015,Li2018,Cheng2018,Yan2018}. And also numerical simulations by \citet{Mei2017} suggested the thickness may exceed $10^3$km.  On the other hand, \citet{Biskamp1993} gave a scaling law for the Taylor scale $l_T$ which represents the inertial-range of the energy spectrum with $l_T=l_{ko}S^{1/6}$, where $l_{ko}$ is the Kolmogorov scale for dissipation and S is the Lundquist number, which indicates that $l_T$ can reach several Mms in the coronal environment. The thickness of the CS and its relation to the reconnection rate
deduced by \citet{Ciaravella2008} could find the theoretical counterpart in \citet{Eyink2013} and
\citet{Lazarian2020}. However, the relation between the CS thickness and the Taylor scale length (see \citealt{Biskamp1993}) is not well understood. 

The fragmented and turbulent CS has been observed in detail by many works \citep{Lin2007,Savage2010,Liu2013,Lin2015,Li2018,Cheng2018,Yan2018,patel2020,Lee2020}. Nonthermal particles observed in the solar eruption suggest the existence of turbulence, and high temperature plasma observed in some events indicates the impact of heating plasmas by turbulence \citep{Warren2018}.

In the work of \citet{Barta2011}, the CS fragmentation and coalescence of plasmoids facilitate the energy release process in the solar flares. \citet{Huang2017} performed a series of 2D simulations of magnetic reconnection in the evolving CS. They find that the classical Spitzer resistivity is important only in a narrow layer near the resonant surface inside the CS during the linear phase of the tearing mode. This layer is also known as the resistive layer (e.g., see also \citealt{Biskamp1993}). The growth of the tearing mode is associated with the development of plasmoids in both size and number. As the plasmoid becomes wider than the narrow layer, the electric current density increases apparently, and gets oscillating violently (e.g., see also \citealt{shen2011}). At this time, the initial integrity of the CS breaks down and the fast reconnection phase starts. 

Two-dimensional numerical experiments of high resolution of \citet{Dong2018} revealed that the index of the energy spectrum is about $-1.5$ in the inertial stage, and the copious formation of plasmoids results in a sub-inertial range with a spectrum index of $-2.2$.  Many dissipation sites are distributed all over the large-scale CS, and the diffusion in the CS is significantly enhanced, which is equivalent to adding an extra diffusivity in the reconnection region, as suggested by \citet{Lin2007} and \citet{Lin2009}. In the work of \citet{Ye2019}, three types of turbulence were recognized in the CS that is located between the CME and the associated flare. Their 2.5D simulation indicated that the turbulence inside the CS shows the anisotropicity and that on the flare loop top is roughly isotropic.

According to these works and on the basis of our previous works, we are looking into details of magnetic reconnection in the CS above the two ribbon flare (see Figure 1 of \citealt{Kopp1976} and/or Figure 1 of \citealt{Forbes1996}) via 2D simulations. To justify that we adopted 2D model for the actual 3D phenomenon, we argue as follows: Unlike the reconnection process taking place in a 3D homogeneous framework (e.g., see \citealt{Kowal2017,Kowal2020} and \citealt{Beresnyak2017}), the reconnection process taking place above the two-ribbon flare is highly confined to a plate-like CS, so it is an inhomogeneous process.

Both theories (\citealt{Lin2000} and \citet{Lin2002b}) and observations (\citealt{Ko2003} and \citealt{Lin2005}) indicated that the solar eruption is initiated by the loss of equilibrium in the coronal magnetic configuration, and leads to thrusting the upper part of the configuration and stretching the lower part (refer to Figure 1 of \citet{Forbes2000}). The disrupting magnetic configuration usually includes an electric-current carrying flux rope, which is used to model the prominence or filament that floats in the corona. Stretching the lower part of the configuration results in the formation of the current sheet between two magnetic fields of opposite polarity, and thrusting the upper part of the configuration (flux rope) produced an area of low pressure around the current sheet (see Figure 1 of \citealt{Lin2005}). The difference in the pressure between the region near the current sheet and that far from the current sheet pushes both magnetic field and plasma to flow toward the current sheet, constituting the reconnection inflow (see blue arrows in Figure 1 of \citealt{Lin2005}) and invoking the so-called driven reconnection in the plate-like current sheet.

Therefore, the reconnection process that we are studying here is occurring in a region that is highly squeezed in one direction by the reconnection inflow. This yields two consequences: First, magnetic reconnection basically takes place roughly in a 2D space; second, the process occurring in this fashion is inhomogeneous since the freedom of the process in one direction is limited. We note here that the limit to the freedom is not due to the existence of magnetic field, but due to the reconnection inflow. Hence, the reconnection process occurring in the CS above the two-ribbon flare is both driven and inhomogeneous, which is different from that studied by \citet{Kowal2017,Kowal2020} and \citet{Beresnyak2017}. This is why 2D simulations could be used for the actual 3D phenomenon of our interest.

\citet{Lazarian2020} also classified reconnection into 2D and 3D such that the tearing reconnection dominates in 2D while turbulent reconnection process plays a key role in 3D cases. Looking into details of the reconnection processes of the two kinds, we realize that the 2D process dominated by the tearing mode is actually of the inhomogeneous turbulence, and that of the 3D process dominated by the turbulence is in fact of the homogeneous turbulence according to \citet{Biskamp1993}. Numerical experiments also show that the fine structures seen in the planar cuts of 3D CS based on \citet{Titov1999} model are very similar to 2D simulations \citep{Mei2017,Ye2019}.

For the large-scale process in the early stage of reconnection occurring in the coronal environment as presented here, the frozen-in condition is only violated at places where reconnection occurs as discussed by \citet{Eyink2015}, and the scenario of the energy conversion in the CME-flare CS in the 2D fashion could still exist in reality \citep{Guo2015, Yang2020,Lazarian2019,Lazarian2020}. Hence, the reconnection process in 2D and 2.5D occurring in the CME-flare CS as a result of the tearing mode for the onset of fast reconnection is worth looking into as well.

In this work, we perform a 2D numerical study for magnetic reconnection in the CS occurring in the classical two-ribbon flare model \citep{Petschek1964,Carmichael1964,Sturrock1966,Hirayama1974,Kopp1976,Lin1995,Lin2004}. In next section, we introduce the model and the code used in this study. Section \ref{sec:results} gives the numerical results and the related analyses for reconnection, and properties of the associated turbulence. Finally, we summarize the work in Section \ref{sec:summary}.

\section{Numerical models and methods}
\label{sec:method}

This work focuses on the CS above the two-ribbon flare given by the CSHKP model \citep{Kopp1976}. Our simulation starts with a configuration in equilibrium, which includes two magnetic fields of opposite polarity perpendicular to the bottom boundary that is located on the photospheric surface. The governing MHD equations including the gravity and resistivity read as:

\begin{equation}
   \frac {\partial \rho  }{ \partial t } +\nabla \cdot \left( \rho \mathbf {v} \right) =0,
\end{equation}
\begin{equation}
    \frac { \partial \mathbf{B} }{ \partial t } =\nabla \times \left( \mathbf{v}\times \mathbf{B}\right) +\frac {1 }{ S} \nabla ^2 \mathbf{B},
    \label{eq:magnetic induction}
\end{equation}
\begin{equation}
    \rho\left[ \frac { \partial \mathbf{v} }{ \partial t }+ \left( \mathbf{v}\cdot \nabla  \right)\mathbf{v}  \right] =-\nabla p+\mathbf{J}\times \mathbf{B}+\rho \mathbf{g},
\end{equation}
\begin{equation}
    \frac { \partial e }{ \partial t } +\nabla \cdot \left[ \left( e+P' \right)\mathbf{v} - \left( \mathbf{v}\cdot \mathbf{B} \right) \mathbf{B} \right] =\rho \mathbf{g}\cdot \mathbf{v} +\frac { 1 }{ S } \nabla \cdot \left[\mathbf{B} \times(\nabla \times \mathbf{B} )\right],
\end{equation}
\begin{equation}
    p = \rho T,
    \label{eq:EOS}
\end{equation}
\begin{equation}
    \nabla \cdot \bf{B}=0.
\end{equation}

Here, all the physical quantities are dimensionless. They are almost duplicated from that of \citet{shen2011}. The quantities $\rho$, $\bf{v}$, $\bf{B}$, $p$, $\bf{J}$, $T$ are mass density, velocity, magnetic field, gas pressure, current density and temperature, respectively. The energy density $e={ \rho v^2 }/{ 2 }+{p  }/{ (\gamma-1) }+{B^2  }/{2  }$ while $\gamma$ (set to $5/3$ for the ideal gas) is the ratio of the specific heat, $\bf{g}$ is the gravity, $P'=P+{B^2  }/{ 2 }$ is the total pressure including the gas pressure and the magnetic pressure, $S=L_0V_A/ \eta$ is the Lundquist number, where $L_0$ is the characteristic length, $v_A$ is the Alfv\'en speed and $\eta$ is the magnetic diffusivity. In our simulations, the characteristic values are $B_0 = 0.01T$, $L_0 = 10^8$m, $\rho_0 = 1.67\times 10^{-12} $kg/m$^3$. Given these values, we obtain 
$v_A=6.9\times 10^5$m/s, $t_0= 14.5$s, $P_0 = 80$Pa, $J_0=7.95\times 10^{-5}$A/m$^2$ and $T_0 = 2.90 \times 10^9$K as the characteristic values for velocity, time, gas pressure, current density and temperature.

The dimensionless gravity reads as
\begin{equation}
    \mathbf{g}=-\frac { GM_{ \odot} }{ {(y\cdot L_{ 0 }+R_{ \odot })}^2 } \cdot \frac { \rho_0 L_0}{P_0 } \hat { y } ,
    \label{eq:grav}
\end{equation}
with $M_\odot = 1.99\times10^{30} $kg, $R_\odot =6.96\times 10^8$m ,$G=6.672\times10^{-11}$N$\cdot$m$^2$/kg$^2$ being the mass of the Sun, the radius of the Sun and the gravitational constant; and $\hat{y}$ being the unit vector in the $y$-direction. We set $g_m={GM_ \odot \rho_0 L_0 }/{ P_0 }$ to make Eq. \ref{eq:grav} more concise and it becomes $\mathbf{g} = { g_m }\hat{y}/{ \left(y \cdot L_0+R_{\odot} \right)^2  } $.

Regarding the initial conditions, we construct a Harris-like current sheet described by:
\begin{equation}
    B_{x} = 0,
\end{equation}
\begin{equation}
B_{y}=
\left\{
\begin{array}{rcl}
\sin {\left(  {\pi x  }/{ 2w  }\right)  } ,     &      & \left| x \right|  \le w,\\ 
1, &      & x>w , \\
-1,&    & x<w,
\end{array} \right.
\label{eq:by}
\end{equation}
\begin{equation}
    B_z = 0.
\end{equation}
The background magnetic field $B_y$ in our simulation follows the typical sine-type current sheet which follows the work by \citet{Forbes1983}, \citet{Forbes1991}, \citet{shen2011,shen2013} and \citet{Ye2020} with $w$ in the Eq. (\ref{eq:by}) being the half-width of CS and is set to be 0.1 initially. To initiate the evolution in the system, we add a small perturbation to the initial configuration at point (0,$y_c$) defined as:
\begin{equation}
    A_{\epsilon }=\epsilon \exp
\left[-\left(\frac{x}{l_{x}}\right)^{2}-
\left(\frac{y-y_{c}}{l_{c}}\right)^{2}
\right] ,
    \label{eq:a_pert}
\end{equation}
where $\epsilon=0.03$, $l_x=0.01$, $l_y=0.01$ and $y_c=0.5$ are the amplitude of the perturbation, dimensionless perturbation wavelengths in $x$-and $y$-directions and the location where the perturbation occurs, respectively.

The initial temperature and pressure distributions are set as below:
\begin{equation}
    T(y) = \frac { T_{cor}+T_{chr} }{ 2 }+\frac { T_{cor}-T_{chr} }{ 2 } \times \tanh{\left[ \left( y-h \right)/{\theta  } \right] } ,
\end{equation}
\begin{equation}
   P(y) = P_{cor} {\rm exp} [{\frac{g_m }{T_{cor}L_0}(\frac{1}{yL_0+R_\odot}-\frac{1}{h_{c}L_0+R_\odot}})] , y\ge h+10\theta  ,
\end{equation}
\begin{equation}
   P(y) = P_{chr} {\rm exp}[-\int_{0}^{y}\frac{g_m}{ T(y)}{(yL_0+R_\odot)^{-2}}dy] , y< h+10\theta,
   \label{eq:p at chr}
\end{equation}
where
\begin{equation}
 P_{chr} = P_{cor}{\rm exp}[\int_{0}^{h+10\theta}\frac{g_m}{T(y)}{(yL_0+R_\odot)^{-2}}dy]  .
\end{equation}
In above equations, $T_{cor}=6.90\times 10^{-4}$ and $T_{chr}=1.90\times 10^{-6}$ are the dimensionless temperatures for the corona and the chromosphere, respectively. For $y< h+10\theta $, the chromosphere is located, $y \ge h+10\theta$ is for the corona, $h=0.03$ and $\theta=0.003$ are the height and the width of the transition region; $P_{cor}=0.01$ is the gas pressure of the corona. The gravitationally stratified atmosphere consists of two parts, and the density distribution in the simulation domain is given by Eq. (\ref{eq:EOS}) and $\rho = {p}/{T}$.

As for boundary conditions, we set the line-tied boundary at the bottom $y=0$, and the open boundary for the other three sides, through which plasma can enter or exit freely. Following \citet{shen2011}, we have the magnetic field:
\begin{equation}
    \frac { \partial B_y\left(x,y=0,t  \right)  }{ \partial t } =0.
    \label{eq:line-tiedb}
\end{equation}
To prevent the plasma at the bottom from slipping, we have:
\begin{equation}
    \mathbf{v}\left(x,y=0,t  \right) =0,
\end{equation}
and for the mass conservation on the bottom, we have:
\begin{equation}
    \frac { \partial \rho\left( x,y=0,t \right)  }{ \partial y  }  =0 ,
    \text{and  } \frac { \partial p\left( x,y=0,t \right)  }{ \partial y  }  =0.
\end{equation}

The simulation is performed using ATHENA code v4.2 developed by \citet{Stone2008}. We first performed our simulations under three grid resolutions of $1920\times1920$, $3840\times3840$ and $7680\times7680$ to look into the impact of the numerical diffusion on the physical scenario. The results suggest that the impact in the case of $1920\times1920$ is too apparent to allow the behavior of the system to match the setup of the Lundquist number, say $S=10^6$. \citet{Ye2020}
pointed out that the numerical diffusion due to the low grid resolution may suppress the effective Lundquist number. Therefore, we choose the results corresponding to the high grid resolution to perform the further studies in the work below.

\section{Simulation results}
\label{sec:results}
\subsection{Global Evolution}
\begin{figure*}
\centering
\includegraphics[width=7in]{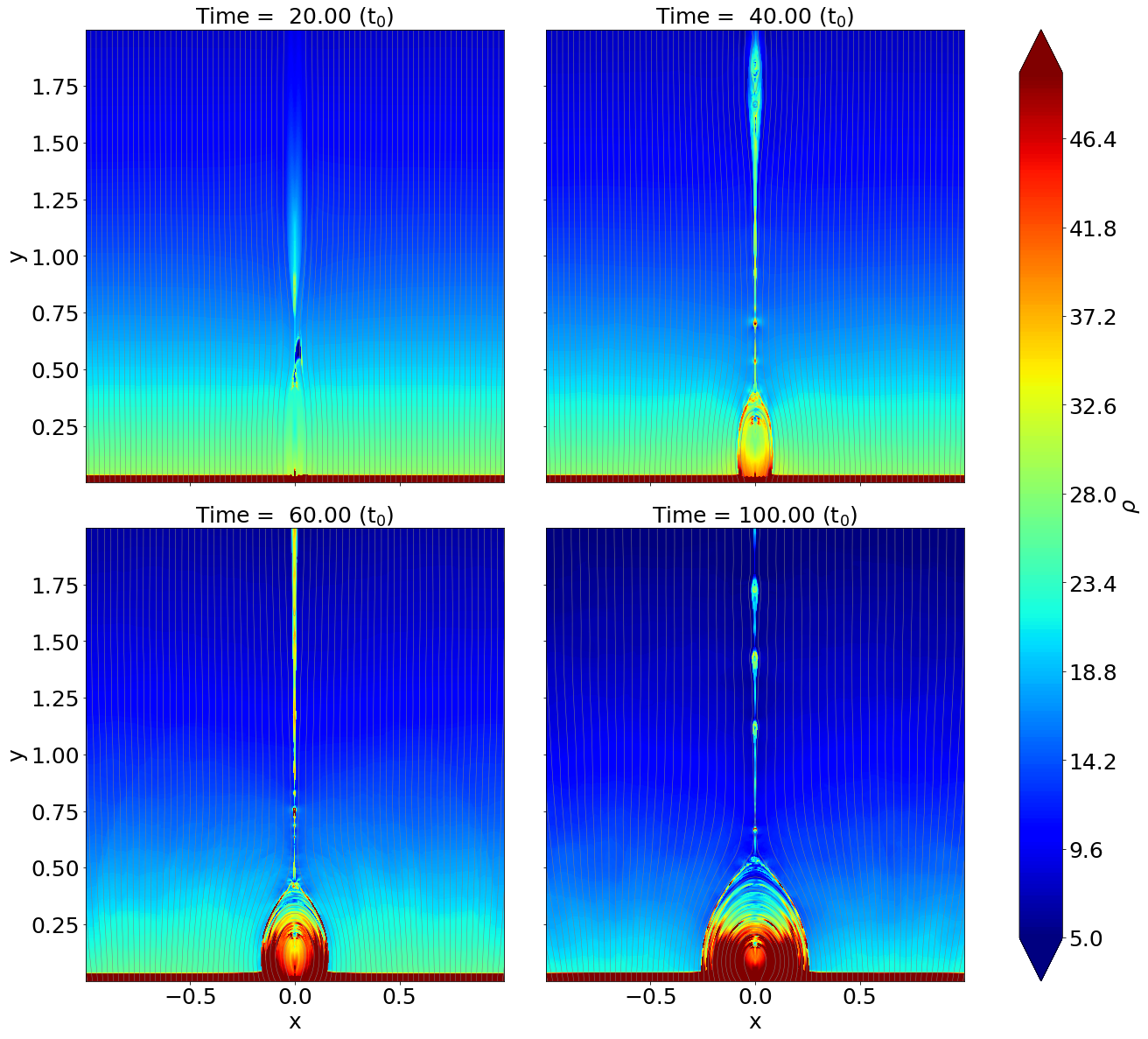}
\caption{Snapshots of the density distribution at time $t=20$, 40, 60 and 100. The gray lines describe the magnetic field at different times.}
\label{fig:total evo}
\end{figure*}

The global evolution in the CS is displayed in Fig.\ref{fig:total evo} which shows mass density distribution in the time interval from $t=20$ to $t=100$.
The simulation starts with the current sheet being squeezed quickly near a specific point and the flare loop begins to appear at the bottom. As the CS becomes thin enough, the tearing mode instability takes place and many plasmoids are produced with multiple X-points occurring between every pair of plasmoids. In this process, that specific point eventually evolves to an X-point at which magnetic reconnection always undergoes faster than at any other X-points. This special X-point is defined as the principal X-point (PX-point). At $t=40$ the first plasmoid appears in the CS, and the reconnection enters the impulsive phase with more plasmoids appearing and moving bidirectionally. Some of them fall and collide with flare loops and finally form a dense shell of flare loops, while the others move upwards and flow out of the upper boundary. Later at $t=60$, a low-density cavity is formed above the flare loop. At $t=100$, the bidirectional moving plasmoids are clearly seen in the reconnection outflows. 

\begin{figure}
    \centering
    \includegraphics[width=3in]{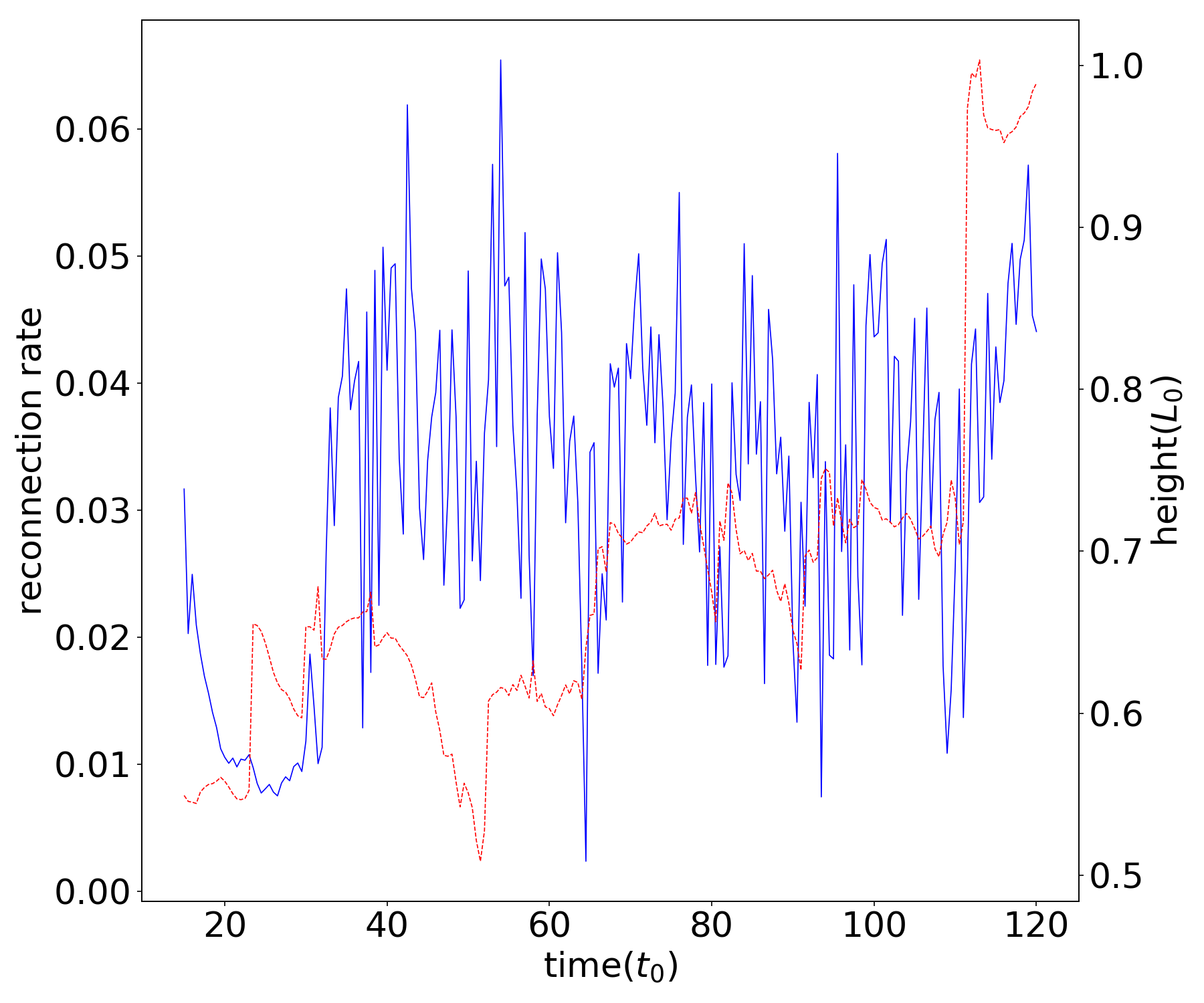}
    \caption{Reconnection rate and PX-point height in the simulation with grid resolution of $N_g=7680 \times 7680$. The blue solid line represents the evolution of reconnection rate. The red dashed line shows the height of P-X point in the simulation.}
    \label{fig:pxheight}
\end{figure}

Motions of the PX-point shown in Fig. \ref{fig:pxheight} display a very different feature from those shown by \citet{shen2011}, which indicated that the PX-point moves upward with a small amplitude oscillation around the stagnation point (S-point), and the reconnection outflow right behind the plasmoid moves faster than this plasmoid. Fig. \ref{fig:pxheight} displays that the PX-point moves in the similar fashion at the beginning until $t=40$ when it starts moving downward, and manifests a jump at about $t=50$. The same pattern repeats at $t=90$ and $t=110$, respectively. Looking carefully at the reconnection process and the motion of plasmoids created in this process, we realized that the gravity plays an important role in the kinematic behavior of plasmoids.

\citet{shen2013} pointed out that a plasmoid continues to grow in both mass and volume after formation as magnetic reconnection progresses. In the case of the gravity absent, the motion of the plasmoid is not affected by the mass accumulation; when the impact of the gravity is included, on the other hand, the situation changes. With the continuous increase in mass, the impact of the gravity on the plasmoid motion gets more and more apparent. As the initial kinetic energy possessed by the upward plasmoid after leaving the PX-point is totally converted into the gravitation potential energy, and the reconnection outflow behind is unable to push the plasmoid to move upward furthermore, the plasmoid will turn to move downward. This forces the PX-point and the other plasmoids below to fall together and eventually merge with the flare loop. The previous PX-point disappears and the associated magnetic structure is destroyed as well. Meanwhile an ordinary X-point above the heavy plasmoid automatically upgrades to the new PX-point. This process happens very quickly, almost at the same time as the previous PX-point disappears, the new PX-point is determined. Thus we see from Fig. \ref{fig:pxheight} that a jump in the PX-point height occurs following a gradual descent of the height. As for which ordinary X-point upgrades to the new PX-point, it is an open question, and we shall investigate it further in the future.

We then evaluate the reconnection rate near the PX-point in the way of: $M_A={v_{in}  }/{ v_A } $ where $v_{in}$ and $v_A$ are the inflow velocity and the local Alfv\'en velocity near the PX-point, respectively.
As shown by Fig.\ref{fig:pxheight}, the reconnection 
goes slowly at the beginning of the simulation. As the tearing mode instability is invoked in the CS, the process turns into fast reconnection phase, and the reconnection rate jumps from $0.01$ to $0.04-0.06$. 

\subsection{Numerical Diffusion and Extra Dissipation}
In our simulation, the Spitzer resistivity is set to be $10^{-6}$. Of course the numerical diffusion is inevitable. The numerical diffusion enhances the dissipation in the fluid, decreases the effective Lundquist number, and thus it suppresses the occurrence of the tearing mode instability. \citet{shen2011} used the AMR-improved SHASTA code with the grid size of 333~km to study the fine structure in the CS and found that the numerical diffusion brings about 20$\%$ error into the calculation. \citet{Mei2012} studied the eruption of a magnetic flux rope applying NIRVANA code with the grid size of 2000~km and reported that the numerical diffusion ranges from 10 to 20$\%$ of the physical diffusion. \citet{Ye2019} also used NIRVANA code to study the energy cascading in the CS with the smallest grid size of 7.5~km. They showed that the equivalent numerical diffusivity starts from 12$\%$ at the beginning, drastically falls to 4$\%$ and tends to be flat around 2$\%$ once AMR is turned on. They found that for the case of the Lundquist number $S=10^6$, the resolution of $3840\times3840$ could apparently suppress the numerical error and allow the effective Lundquist number to match the prerequisite one.

For the physical scenario manifested by the system we are investigating, the numerical diffusion is considered extra in addition to the classical (or Spitzer) diffusion. Here using the term ``extra'' implies that the numerical diffusion itself is not the only issue that may impact the reconnection process, and that the so-called extra diffusion as a result of the turbulence could be another issue that may govern the energy conversion in a more apparent way (e.g., see \citealt{Lin2015,Ni2018,Shan2021}). Following the practice of \citet{Shan2021}, we study the extra diffusion by looking the ratio given below:
\begin{equation}
    \frac {\eta_n  }{ \eta_m } =\frac { \left| \partial_t A-v\times B +\eta_m\nabla \times B \right|  }{ \left| \eta_m \nabla \times B \right|   }   \text{,}
    \label{eq:difratio}
\end{equation}
where $\eta_n$ represents the extra diffusivity, $\eta_m$ represents the Spitzer resistivity and $\mathbf{A}$ is the associated magnetic potential vector. We note here that the ratio in Eq. \ref{eq:difratio} is evaluated in the fashion of average over a region near the PX-point in order to suppress unnecessary errors.

In addition, we note here that, in principle, the impact of the numerical diffusion on the reconnection process could be calculated via the induction equation directly. We point out that, on the other hand, since the second order differentiation is involved in the calculation and more extra error could be introduced if the induction equation is directly used, we choose to evaluate the impact of the numerical diffusion via Eq. (\ref{eq:difratio}) instead. Although Eq. (\ref{eq:difratio}) here has the same form as that of \citet{Mei2012}, it possesses different meaning here.

\begin{figure}
    \centering
    \includegraphics[width=3in]{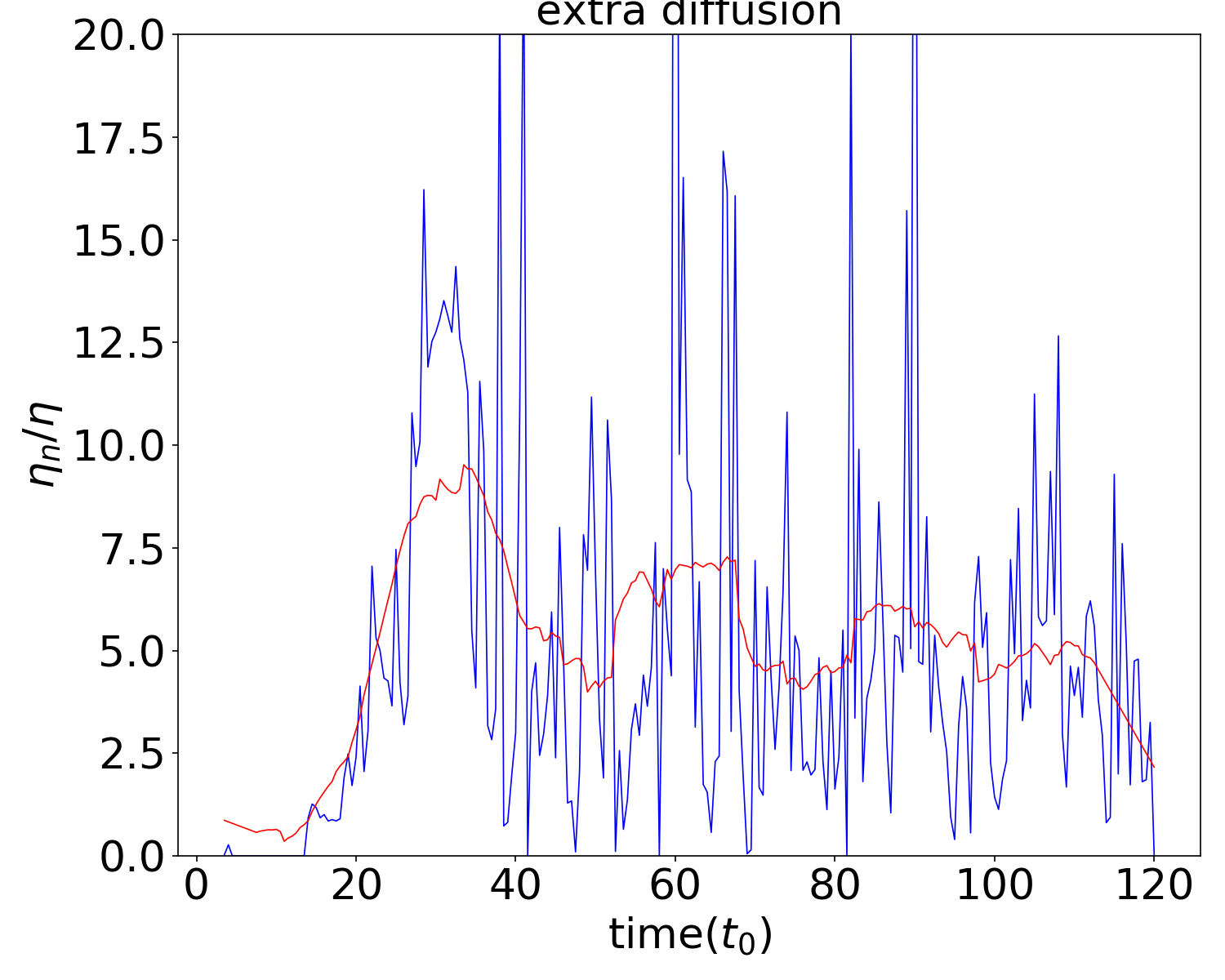}
    \caption{Ratio of extra diffusion in the simulation to Ohmic diffusivity with time. The blue line shows the primitive ratio calculated in the numerical simulation. And the red line is the average ratio.}
    \label{fig:eta}
\end{figure}

To evaluate this ratio, we use the "Userwork-in-loop" block in the ATHENA code \citep{Stone2008} to compute it at each timestep in simulations. This calculation can effectively improve the accuracy compared to the calculation outside the loop, and the ratio in our simulation is shown in Fig. \ref{fig:eta}. 

In principle, the numerical diffusion itself for a given algorithm and the associated grid resolution is roughly fixed. In the initial stage of the simulation, the reconnection process goes very slowly and the ratio is about $0.2-0.3$ as shown in Fig. \ref{fig:eta} that is consistent with the result of \citet{Shan2021}, and could be ascribed to the numerical diffusion. With the appearance of the plasmoid in the CS, the ratio gets big dramatically. Consequently, a lot of plasmoids are formed, which suggests the occurrence of the tearing mode \citep{Furth1963}. The ratio jumps to the range from 5-10 correspondingly. This implies that the extra diffusion becomes dominated by another dissipation term as a result of the fast reconnection phase as indicated by \citet{Eyink2011} and \citet{Lazarian2020}. However, we should note here that fast reconnection accelerates the dissipation of the magnetic field, and the magnetic energy is mainly converted in the kinetic energy in this process, which is basically different from the resistivity effect which transfers magnetic energy into Ohmic heating (see also \citealt{Eyink2011,Lazarian2020}).

\subsection{Energy Spectra Analysis}

Magnetic reconnection produces several open issues about how energy is transferred from large inertial scale to small dissipation scale. It is widely accepted that this transfer is realized by energy cascading process as a result of turbulence. The tearing mode instability triggers the fragmentation of the large scale CS and invokes the fast energy conversion on the small scale. Our numerical simulation duplicates this process. Usually the energy spectrum for this process possesses a double power-law-like pattern, which demonstrates how the energy cascades from large-scale structure to small-scale ones, and at which scale the Spitzer diffusion starts dominated. This process could be displayed by the distribution of the magnetic energy($E_m$) in the CS versus the wave number of the turbulence. \citet{Barta2011} and \citet{Mallet2017} discussed the power-law distribution of energy in the inertial and dissipative ranges. \citet{Barta2011} investigated the impact of fine structures in the CS on the energy spectrum. Their 2.5D simulation indicated that the fragmented reconnection process yielded the spectral index to be about $-2.14$ in the scale range from 300 to 10,000~km, and the inertial stage of energy cascading ends at about 300~km. Results of \citet{Mallet2017} for the energy spectrum manifested a double power-law fashion, and indicated that the spectral index in the inertial range is between $-{5}/{3}$ and $-2.3$. 

To investigate the energy conversion process, we use fast Fourier transformation to deduce the magnetic energy spectra in the CS during the steady reconnection phase. When the tearing mode instability happens in the CS, plasmoids appear and interact with one another. When plasmoids move upward, some of them will catch up with ones ahead and merge into a bigger one eventually, and the secondary reconnection process takes place during the merging, in which many more smaller fragmented CSs are formed between two merging plasmoids enhancing the magnetic field dissipation. 

\begin{figure}[ht]
\centering
\subfloat{
\begin{minipage}[b]{1.0\textwidth}
\includegraphics[width=6.5in]{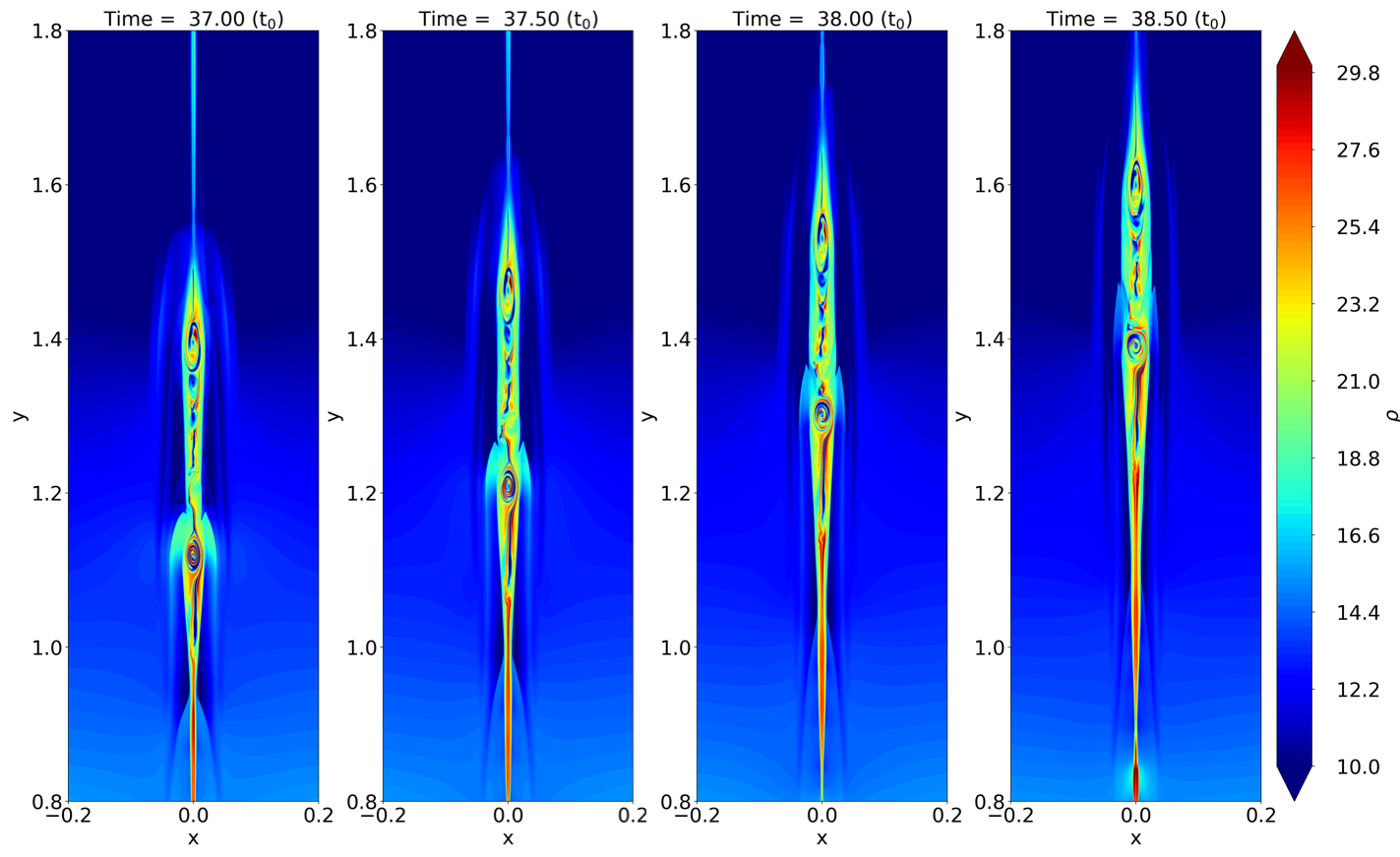}
\end{minipage}
}
\quad
\subfloat{
\begin{minipage}[b]{1.0\textwidth}
\includegraphics[width=6.5in]{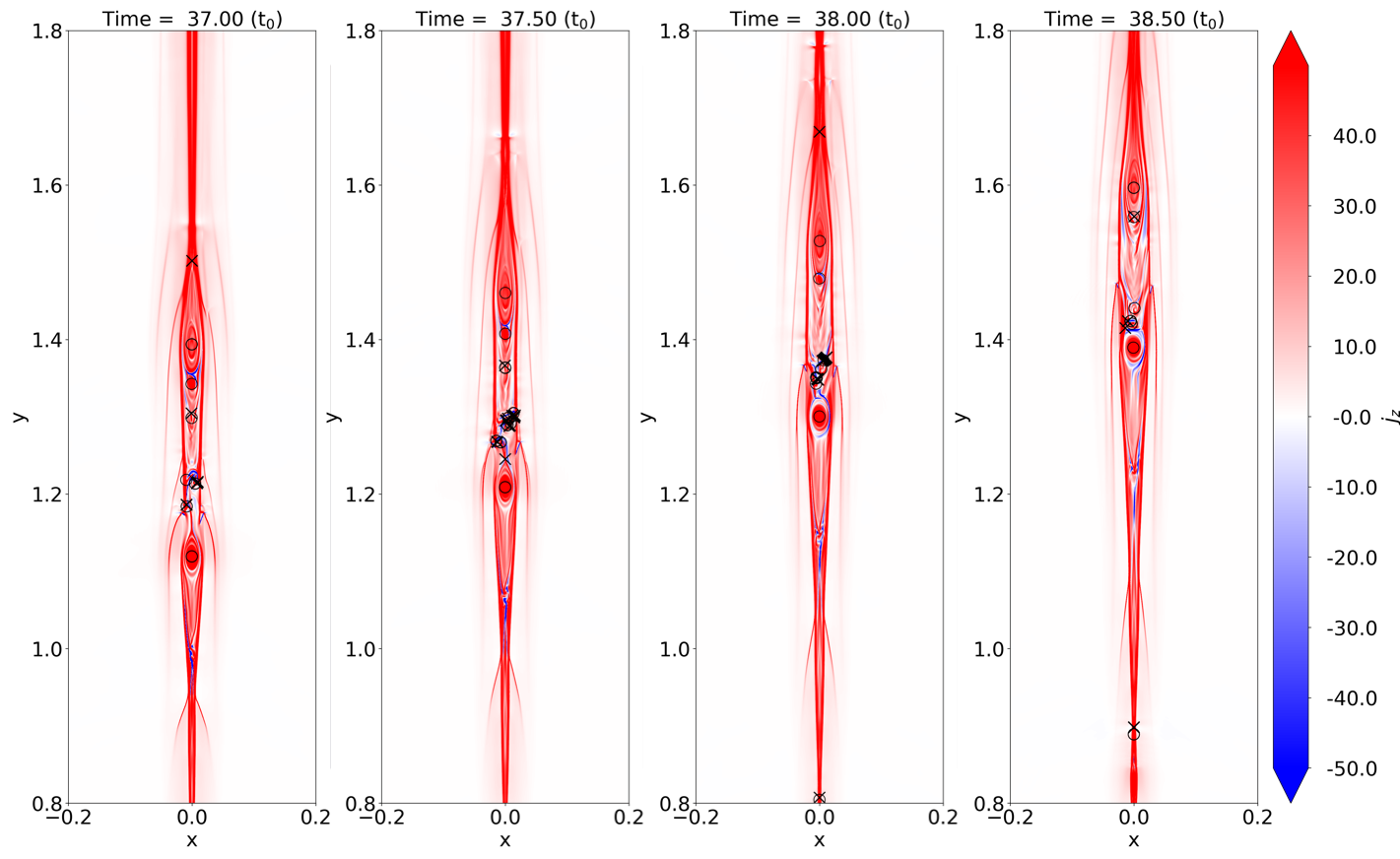}
\end{minipage}
}
\caption{Evolution of density and current density at time of $t=37.0$, $t=37.5$, $t=38.0$ and $t=38.5$. Letters `x' and `o' mark the X-point and the O-point at multiple secondary magnetic reconnection sites.}
\label{fig:collision}
\end{figure}

\begin{figure*}
\centering
\subfloat{
    \begin{minipage}[b]{0.31\textwidth}
    \includegraphics[width=2.0in]{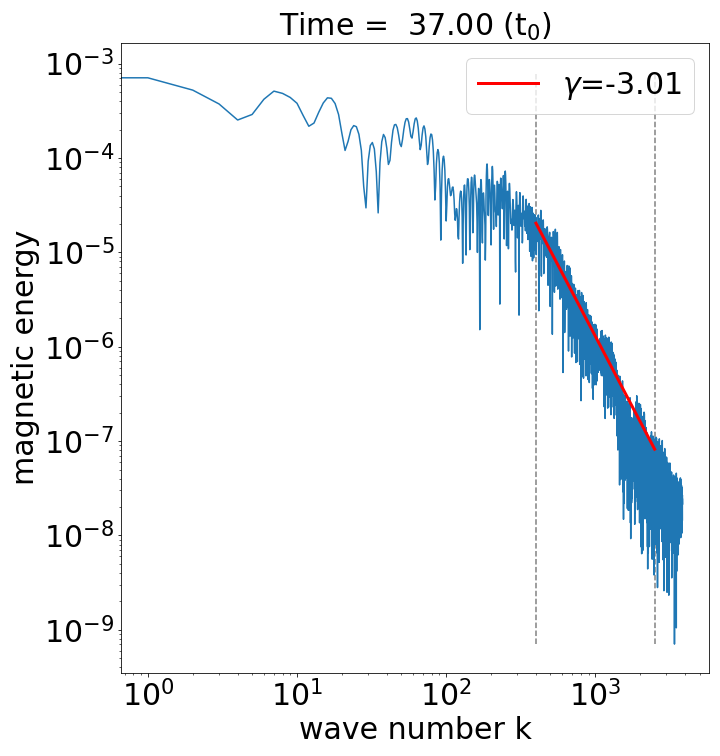}
    \end{minipage}}
\subfloat{
    \begin{minipage}[b]{0.31\textwidth}
    \includegraphics[width=2.0in]{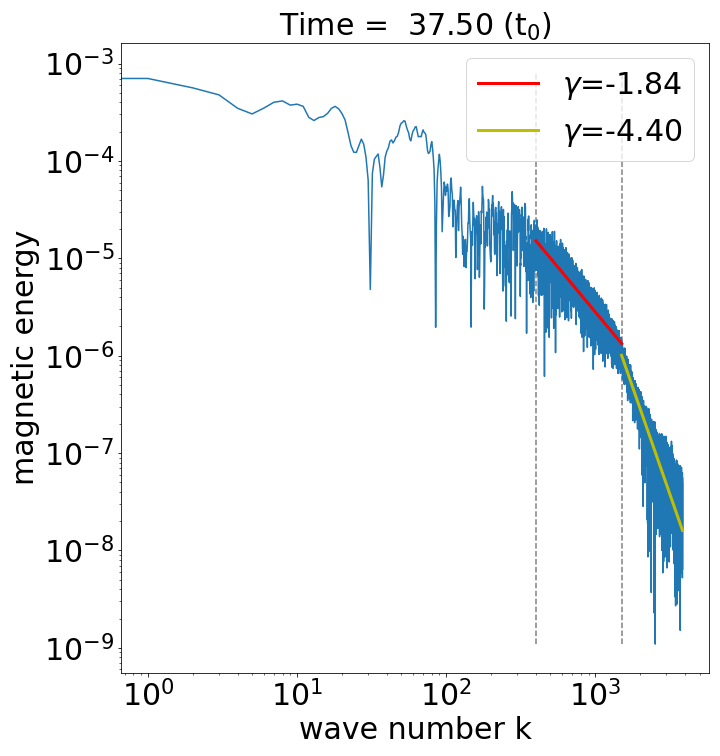}
    \end{minipage}}
\subfloat{
    \begin{minipage}[b]{0.31\textwidth}
    \includegraphics[width=2.0in]{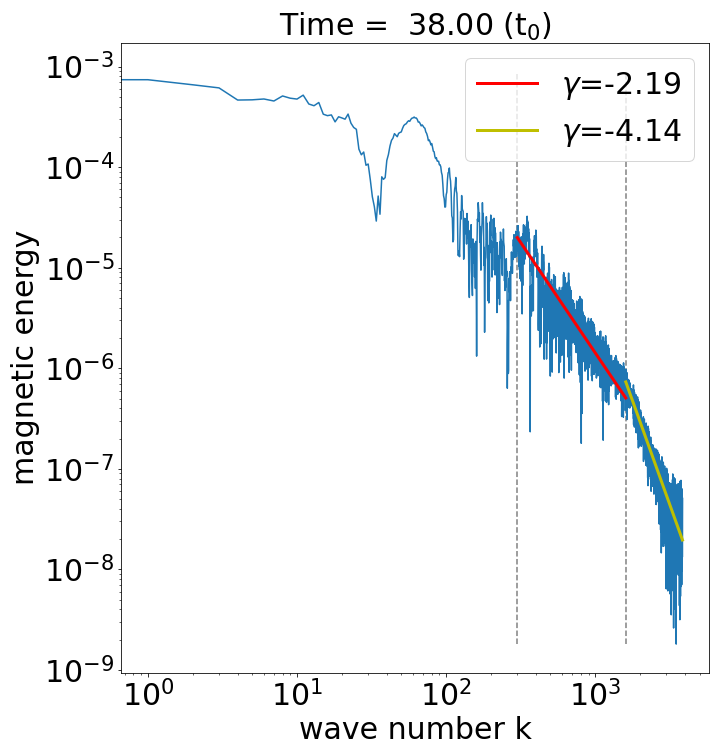}
    \end{minipage}}
\caption{Magnetic energy spectra using 1D FFT technique during a plasmoids collision process. Blue lines are the FFT energy at $t=37.0$, $37.5$ and $38.0$. Red and yellow lines at each time are the fitted power-law distribution of magentic energy. The legends at upper right shows the fitted power indices.}
\label{fig:fft}
\end{figure*}

Fig. \ref{fig:collision} displays the evolution in the CS from $t=37.0$ to $t=38.5$ and shows more details of the secondary small scale structures, as well as their merging. The density distribution between two merging plasmoids looks apparently chaotic and many Sweet-Parker-type CSs appear associated with multiple X-points, which suggests that the diffusion region spread out all over the large-scale CS. A one-dimensional Fourier transform for the magnetic energy distribution along the $y$-axis are performed, and results are given in Fig. \ref{fig:fft}. The power law or double power law distribution pattern can be seen easily, and the corresponding spectral indices are also given.

We notice that before the merging of two plasmoids at $t=37.0$, the energy spectrum presents a single power-law tendency. And the spectral index $\gamma$ is about $-3.01$. When the two plasmoids collide and merge together at $t=37.5$ and $t=38.0$, the magnetic energy spectra show a tendency of double power-law distribution. The turning point of wave number $k$ is at $k\simeq 1,500$ and $k\simeq 1,600$ respectively. The corresponding dissipative scales are about 125~km to 133~km respectively, which are consistent with the width of the fragmented CS appearing between two merging plasmoids whose width is about 192~km. We also calculate more cases and find the width of these fragmented CSs ranges from 100~km to 200~km which is consistent with the scale associated with the turning point in the double power-law spectrum. We further use zero-padding fast Fourier transform (FFT) method to check the energy spectra obtained in the case of the grid resolution $3840 \times 3840$. We find that the turning point does not displace apparently. 

We note here that the scale on which the dissipation becomes dominating in the turbulence is usually believed to be the inertial scale of ions, which is about $10^2$~m in the coronal environment, according to the theory of the classical (namely Spitzer) resistivity. But this scale obtained here is in the range from 100~km to 200~km as indicated in Fig. \ref{fig:fft}. This implies big difference between the expectation of the classical theory and the results here. If the dissipation of the magnetic field occurs through the Spitzer resistivity only, the dissipation scale should stay at a very low level. In reality, on the other hand, the Spitzer resistivity can never be the only dissipative source. For example, the anomalous resistivity due to the ion-acoustic and lower hybrid drift turbulence could produce a dissipative process that is almost 7 orders of magnitude faster than that resulting from the Spitzer resistivity. According to \citet{Strauss1}, the largest scale on which the anomalous resistivity starts being effective is given by: 
\begin{equation}
l_a^\ast =\left( \frac {m_i  }{ m_e }  \right)^{{ 3 }/{ 4 }}\frac { c }{ \omega_{pe}\beta^{{ 1 }/{ 2 } }}  ,
\label{eq:la}
\end{equation}
where $\omega_{pe}$ is the electron plasma frequency and $\beta$ is the plasma $\beta$ in the system of interest, which is 0.1 in this work, and c is the light speed. 

According to the setup for the present simulation, the electron density near the CS is about $10^9$ cm$^{-3}$, which gives $\omega_{pe}=1.78\times 10^8$Hz. Substituting the values of $\omega_{pe}$ and $\beta$ into Eq. (\ref{eq:la}), we have $l_a^\ast=149$~km. Apparently, this scale is large compared to the ion inertial scale in the corona. \citet{Strauss1} pointed out that in the quiet coronal environment, the hyper-resistivity is 9 orders magnitude higher than the anomalous resistivity; and \citet{Lin2007} found that, in the CME/flare CS, the difference is of 4\textasciitilde5 orders of magnitude. The result of \citet{Strauss1} also indicated that both the anomalous and the hyper resistivities depend inversely on the scale of the diffusive structure quadratically. Therefore, in a turbulent CS, the scale $l_h^\ast$ on which the hyper-resistivity tends to dominate diffusion should be related to $l_a^\ast$ and the ratio, $R_{ha}$, of hyper to anomalous resistivities in the way of $l_h^\ast=l_a^{\ast} \sqrt{R_{ha}}$ with $R_{ha}$ ranging from $10^4$ to $10^5$. Thus, we found $l_h^\ast$ ranges from 149~km to 472~km, which is consistent with what we obtained earlier for the dissipative scale deduced from the joint of the double power law spectra.

This indicates that in a turbulent reconnecting current sheet, the Kolmogorov micro-scale could be as large as a few $10^2$~km due to the occurrence of the hyper-resistivity. In the spirit of \citet{Biskamp1993}, we realized that $l_{ko}$ could be somehow related to the thickness, $d$, of the CS in which turbulent magnetic reconncetion is progressing. \citet{Biskamp1993} pointed out that an intermedia spatial scale, the Taylor micro-scale $l_T$, exists between the global scale of the system $L$ and the dissipation scale $l_{ko}$. This means that $L$ cascades to $l_{ko}$ smoothly via $l_T$, and $l_T$ is still located in the inertial range. According to \citet{Biskamp1993}, $l_T=l_{ko}R_m^{1/6}$, so $l_T$ is between $10^3$~km and $2\times 10^3$~km, and $l_{ko}$ between 100~km and 200~km.

Values of $l_T$ deduced here remind us of another important scale in the configuration of magnetic reconnection, namely the thickness of the CS, $d$. Look into the electric current distribution inside the CS along the $x$-direction obtained from our simulations, we notice that the profile of the electric current varies from place to place due to the turbulence in the CS. But the full width of half maximum of the profile is between $1.5\times10^3$~km and $2.5\times10^3$~km, which is consistent with both observations (e.g., see also \citealt{Savage2010,Ciaravella2013,Seaton2017,Yan2018,Cheng2018,Li2018}) and the value of $l_T$ deduced above. This further suggests that the turbulence occurring in the CS greatly speeds up the energy dissipation and allows it to happen at a much larger macro scale, and that the thickness of a turbulent CS should be the Taylor micro-scale of a few $10^3$~km in the coronal circumstance. We also estimate the value of $l_T$ deduced from the results of \citet{Barta2011,shen2013,Ni2015,Ye2019}, and find the consistency with the $l_T$ value obtained in the present work.

\subsection{Width and Area Distribution of Plasmoids}

Copious plasmoids are generated because of the tearing mode instability in the CS and move bidirectionally (Fig. \ref{fig:total evo}). The downward moving plasmoids eventually collide with flare loops and merge into the flare loop system, while the upward moving plasmoids successfully leave simulation domain. \citet{shen2013} investigated the width distribution function of the plasmoids in the CS, and found a power-law distribution in the way of ${ w }^{-2  }$, with $w$ being the width of plasmoid. Following \citet{clauset2009} that gave the power-low distribution via the approach of the maximum likelihood, on the other hand, \citet{patel2020} deduced the distribution function of the plasmoid size as $f(W)\sim{w}^{-1.12}$. 

\begin{figure*}
    \centering
    \includegraphics[width=6.2in]{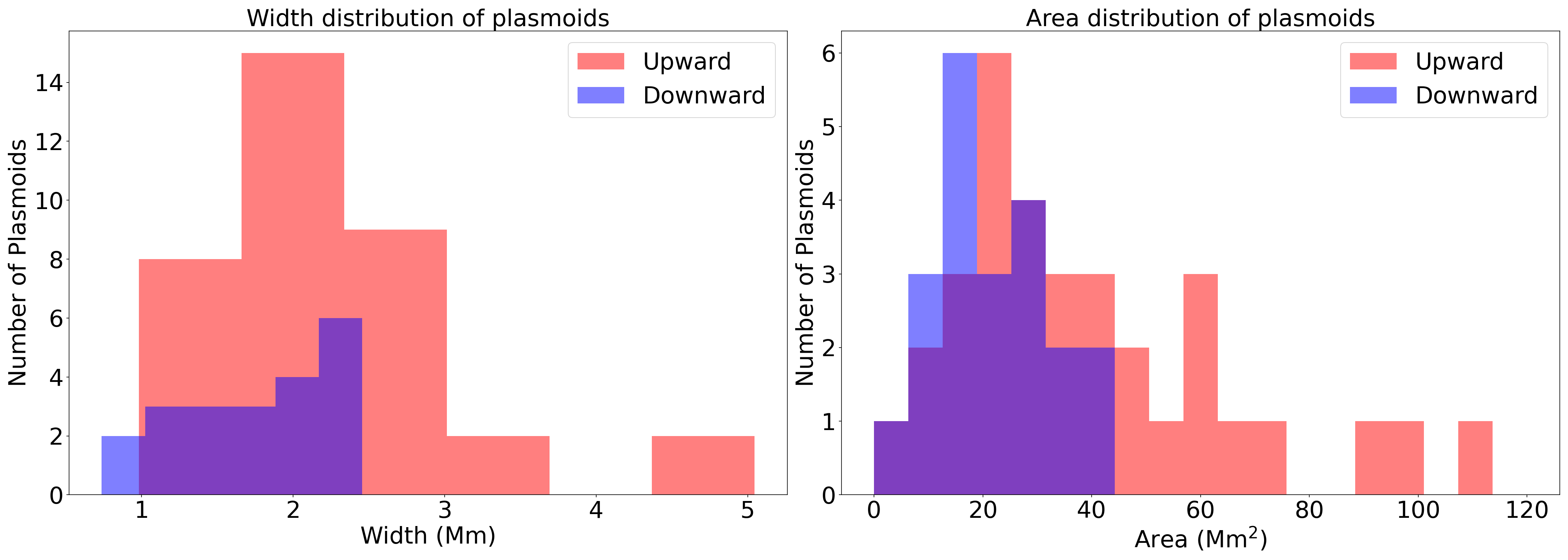}
    \caption{Distribution of plasmoid numbers versus plasmoid width (left) and area (right). The red histograms represent plasmoids moving upward while blue ones represent plasmoids moving downward. }
   \label{fig:width&area}
\end{figure*}

We are able to perform a similar statistical study for our results. We selected 55 plasmoids with 36 moving upward and 19 downward. Distributions of plasmoid number versus width and area are shown in Fig. \ref{fig:width&area}, which indicates that the width of plasmoid could be up to $5\times10^3$~km, while the area up to $10^8$~km$^2$. We noticed that our results are consistent with those of \citet{patel2020} who showed that the width of plasmoids can be up to $10^4$~km and the area can reach up to $8\times10^7$~km$^2$, respectively. 

The average width of these plasmoids in our work is about $2.07\times10^3$ km and the average area is about $4.13\times10^7$ km$^2$, while the median width is about $1.97\times10^3$ km and the median area is about $2.95\times10^7$ km$^2$. Particularly, the sizes of plasmoids moving upward and downward show a little difference. For downward plasmoids, the average width and area are $1.73\times10^3$ km and $2.08\times10^7$ km$^2$ respectively, while for upward ones they are $2.26\times10^3$ km and $5.21\times10^7$ km$^2$. As for median values, the width and area for downward plasmoids are $1.72\times10^3$ km and $1.66\times10^7$ km$^2$ while those for upward ones are $2.09\times10^3$ km and $3.19\times10^7$ km$^2$, respectively. These results are listed in Table \ref{tab:plasmoids}. Usually, both the magnetic and the gas pressure are stronger at the lower altitudes than at the higher altitudes. So the upward moving plasmoids expands more easily and faster than those moving downward, which accounts for the fact that the upward moving plasmoid is fatter than the downward moving one.

\begin{table*}
    \centering
    \begin{tabular}{c c c c c c }
    \hline
        movement & counts & average width($10^3$~km) & average area ($10^7$~km$^2$) & median width ($10^3$~km) & median area ($10^7$~km$^2$)\\
        \hline
        upward &  36 & 2.26 & 5.21 & 2.09 & 3.19\\
        downward & 19 & 1.73 & 2.08 & 1.72 & 1.66 \\
        all & 55 & 2.07 & 4.13 & 1.97 & 2.95  \\
    \hline
    \end{tabular}
    \caption{Statistical Features of Plasmoids Moving Upward and Downward, as well as All Plasmoids.}
    \label{tab:plasmoids}
\end{table*}

\begin{figure*}[ht]
    \centering
    \includegraphics[width=6.2in]{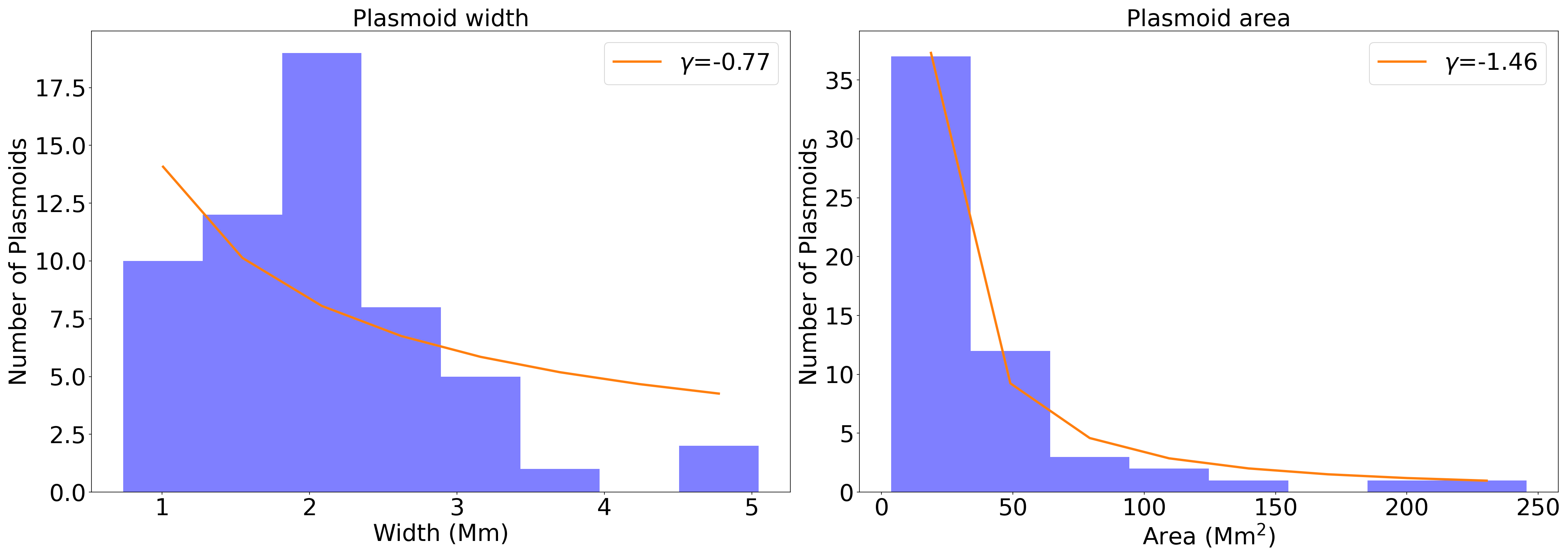}
    \caption{Distributions of numbers of all plasmoids observed moving both upward and downward versus width (left) and area (right). The histograms are for the numbers of plasmoids in each width/area range including both upward and downward plasmoids. The orange lines are the fitting power-law distribution of the width or area. The indices of the power-law distributions are $-0.77$ and $-1.46$ for width and area, respectively. }
\label{fig:plasfit}
\end{figure*}

Furthermore, we plot numbers of all plasmoids observed moving both upward and downward against width and area of the plasmoid in Fig. \ref{fig:plasfit}, in which the left panel is for the number versus width and the right panel for the number versus area. Fitting these two distributions to the power-law function yields the indices of $-0.77$ and $-1.46$, respectively, which are basically consistent with the results of \citet{shen2013} and \citet{patel2020}.

\subsection{Termination Shock and Energy Accumulation Rate}

\begin{figure*}[!htb]
    
    \subfloat[density (colorful shading) and velocity (black arrows)]{
    \begin{minipage}[t]{1.0\textwidth}
    \includegraphics[width=6.5in]{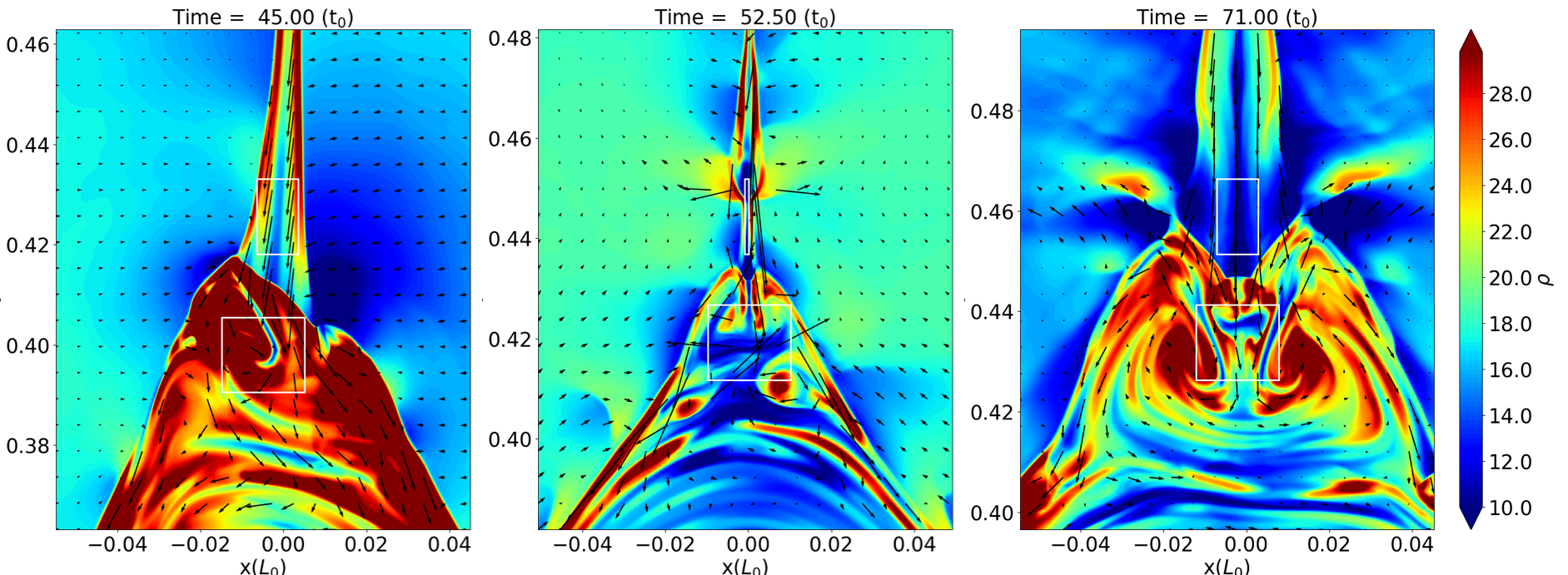}
    \end{minipage}}
    \quad
    \subfloat[Mach number]{
    \begin{minipage}[t]{1.0\textwidth}
    \includegraphics[width=6.5in]{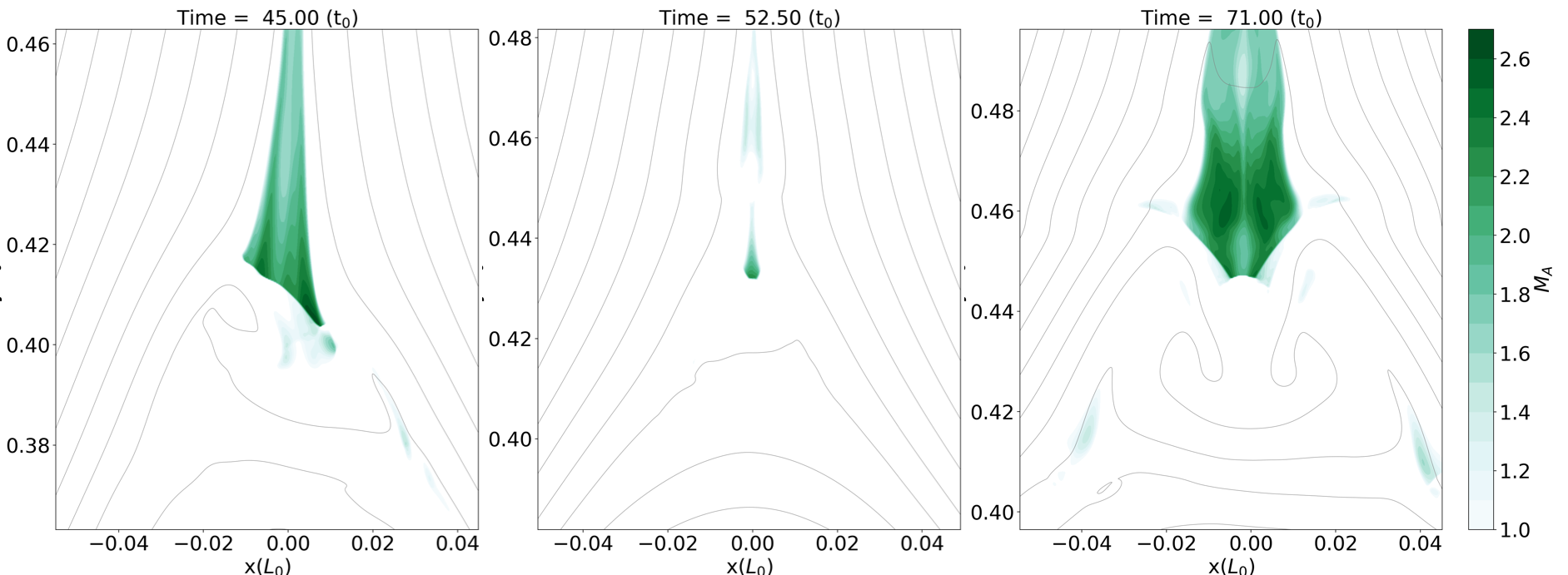}
    \end{minipage}}
    \quad
    \subfloat[plasma $\beta$]{
    \begin{minipage}[t]{1.0\textwidth}
    \includegraphics[width=6.5in]{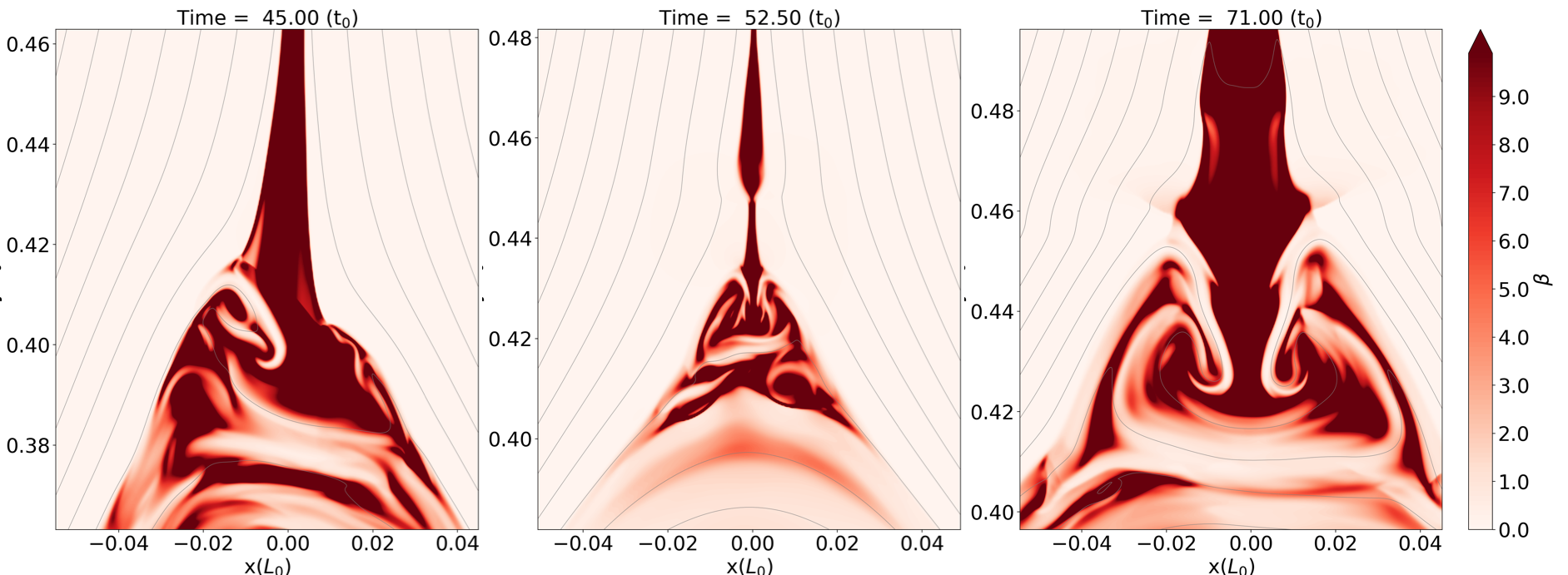}
    \end{minipage}}
    \caption{Distributions of density and velocity (a), Mach number of the downward reconnection outflow (b), and plasma $\beta$ (c) around TS region at $t=45$ (left column), $52.5$ (middle column), and $71$ (right column), respectively. The white rectangle in each panel in row (a) marks the region for evaluating the STD of velocity.}
    \label{fig:tsshape}
\end{figure*}

Termination shock (TS) above the flare loop region is also a topic which attracts much attention in solar physics. It includes many complex structures and plays an important role in energy conversion. It forms between the top of the flare loop and the bottom of the CS. \citet{Forbes1996} pointed out that the termination shock is a result of the interaction of the supersonic reconnection outflow moving downward with the closed flare loop. Fig. \ref{fig:tsshape} shows the distributions of density, Mach number and plasma $\beta$ near the flare loop-top at times $t=45.0$, $52.5$ and $71.0$, respectively. A significant change in the density on both sides of TS and an apparent dividing line which is the shock front could be recognized. The Mach number distribution indicates the supermagnetosonic nature of the reconnection outflow, and it ranges from 1.0 to 2.6. The Mach number of the reconnection outflow before the TS could somehow indicate the energetics of the downflow. We notice that values of the Mach number before TS at above 3 moments are 2.32, 2.30 and 2.42, respectively; and at $t=71.0$, the downflow becomes more energetic, and the corresponding plasma $\beta$ in the related region could reach up to unity. From the density distribution, we obtain the compression ratios across TS at above 3 moments, which are 2.33, 2.63 and 2.67, respectively. More values of the compression ratio at several other times are deduced as well, and the result shows that the compression ratio across the TS is between 2 and 3.

During the whole process, we notice that TSs have different geometries in various stages. Three main shapes are found, including those of linear, V-like and inverse-trapezoid types as shown in Fig. \ref{fig:tsshape}. The linear pattern of TS front, no matter horizontal one or oblique one, mainly results from the interaction between the reconnection outflow and flare loops. While the V-like pattern and inverse-trapezoid pattern are generally the result of the collisions of the plasmoid with the flare loop. 

\begin{figure*}[!htb]
    \centering
    \includegraphics[width=6.2in]{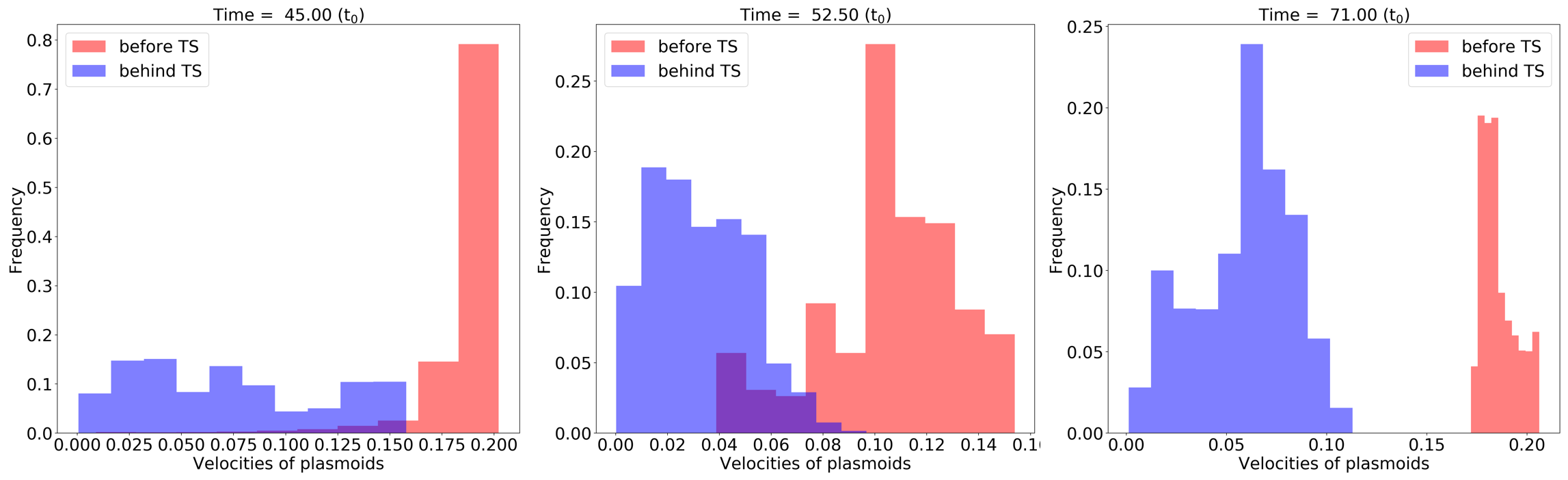}
    \caption{Histograms of the velocity distributions at $t=45$, 52.5 and 71, respectively. The red is for the velocity before TS and the blue one is for that behind TS. The $x$-axis is for the plasmoid velocity, and the $y$-axis is for the normalized frequency of the occurrence of a given velocity.}
    \label{fig:vhisto}
\end{figure*}

To look into how the shape of TS affects the turbulence strength, we use the standard deviation (STD) of velocity as an index of the turbulence strength before and 
behind TS. Generally, the velocity of the downward reconnecting outflow before TS is more uniform than that behind TS. Fig. \ref{fig:vhisto} displays the histogram for the frequency at which a given velocity of the plasma flow occurs either before (red) or behind (blue) the TS. We notice that the distributions of the velocity before TS are usually less dispersive than that behind TS, namely the flow velocity behind the TS spreads in a wide range with large STD. On the other hand, the mean velocities before the TS are apparently higher than those behind the TS, which suggests the occurrence of a sharp deceleration of the plasma flow across the TS.

Comparison of various shapes of TSs shows that the more asymmetric and irregular the TS is, the more turbulent the region behind the TS is. In particular, for the linear TS, the enhancement of the turbulence by the oblique TS is more apparent than the horizontal one. For the oblique TS that is asymmetric, the enhancement factor is between 1.5 and 2; while for the horizontal TS the factor is about 1.0. At $t=45$, the STD before TS is 0.0199 while that behind TS is 0.0449, leading to an enhancement factor of about 2.26. For the regular and symmetric configuration (such as that at $t=52.5$), the STD before and behind TS are just nearly the same, say 0.02. At $t=71.0$ the strengthening of turbulence behind TS is quite apparent with the enhancement factor up to 2.81. Thus irregularity and asymmetry of TS structure are more efficient for enhancing turbulence.

\begin{figure*}[!htb]
    \centering
    \subfloat[energy conversion rate]{
    \begin{minipage}{\textwidth}
    \includegraphics[width=6.2in]{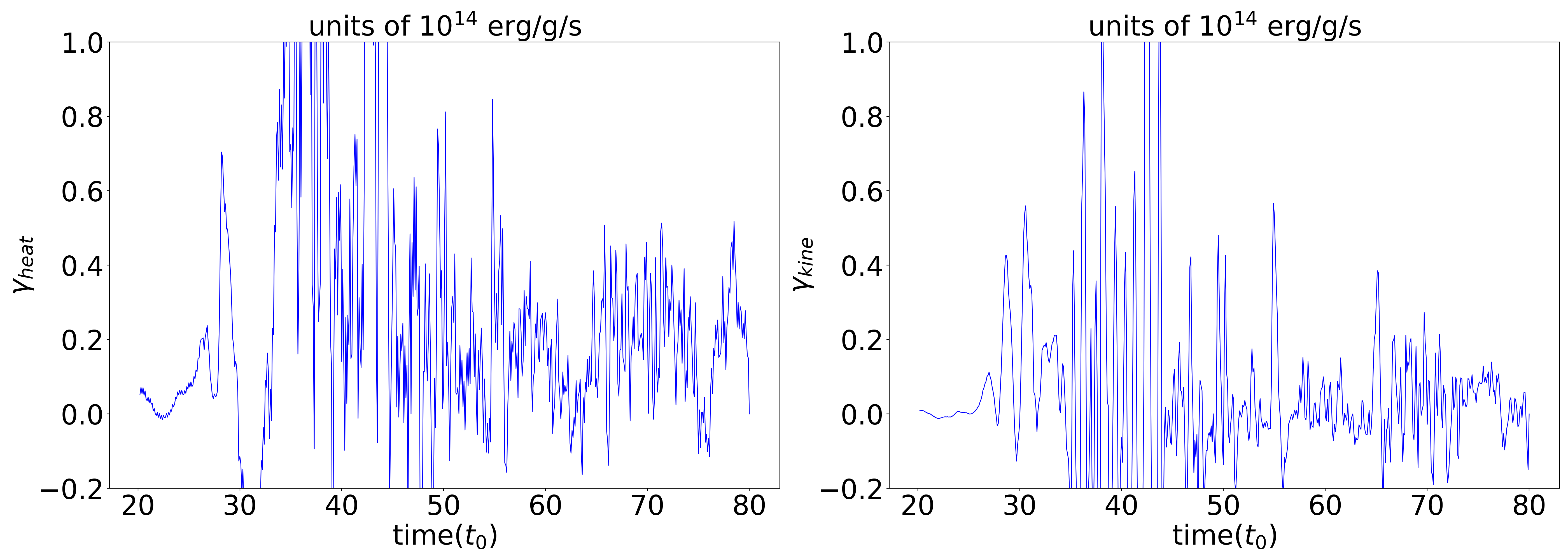}
    \end{minipage}
    \label{fig:ene1}
    }
    \quad
    \subfloat[energy accumulation]{
    \begin{minipage}{\textwidth}
    \centering
    \includegraphics[width=3.1in]{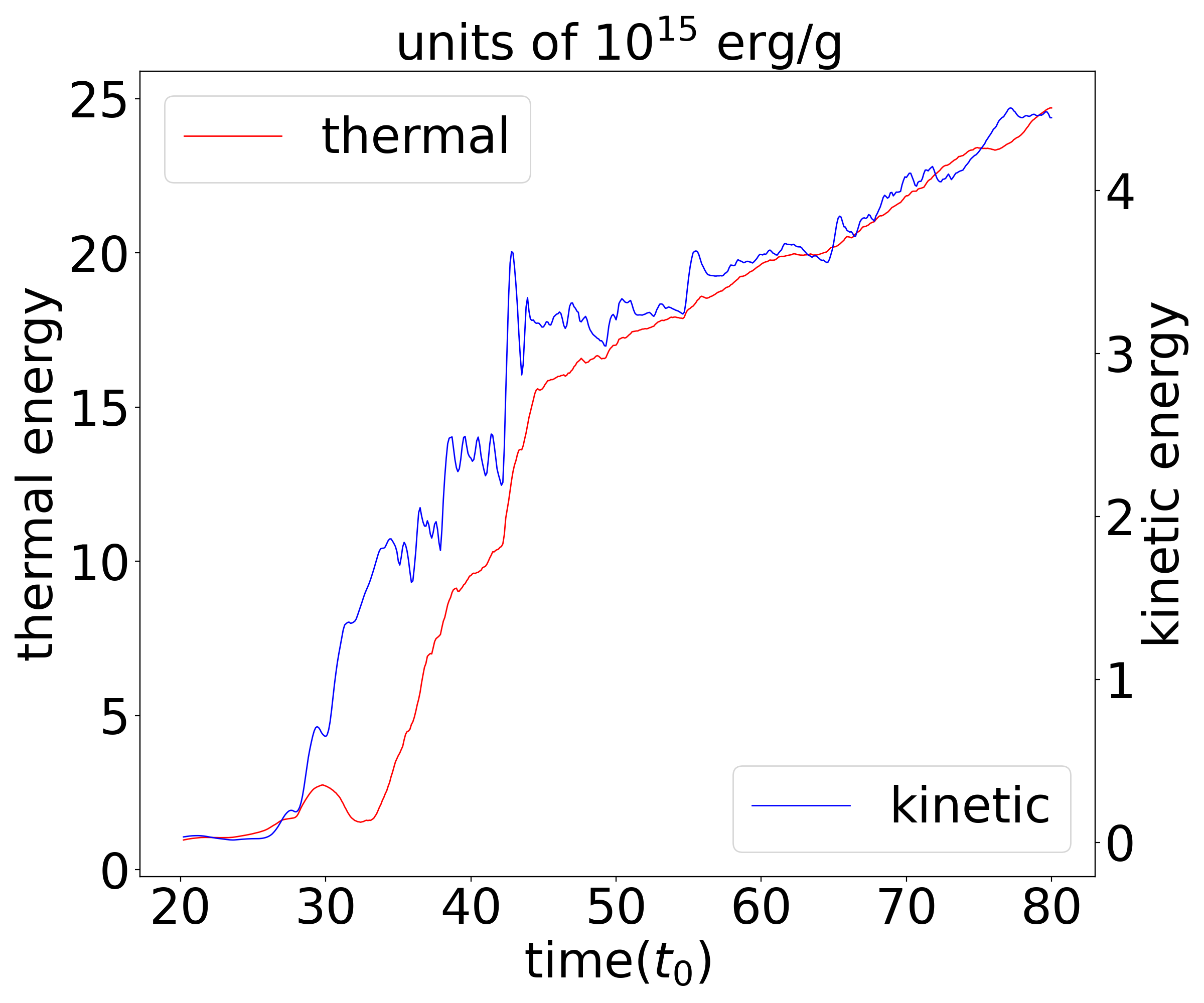}
    \end{minipage}
    \label{fig:ene2}}
    \caption{(a) Energy conversion rates for thermal energy (up panel) and kinetic energy (down panel) with time. (b) Accumulative thermal and kinetic energy transfer rate from time $t=20$ to 80. The red curve is the accumulation of thermal energy while the blue one represents kinetic energy.}
\label{fig:transferrate}
\end{figure*}

To study the energy conversion efficiency in the region around TS, we evaluate the kinetic energy and the thermal energy. We first locate the TS position by calculating $\nabla \cdot \bf { v } $. Due to the symmetry about the $y$-axis, the center of TS is very close to $x=0$. We select $\Omega$ of $[x_{TS}(t)-0.05$, $x_{TS}(t)+0.05]\times [y_{TS}(t)-0.05$, $y_{TS}(t)+0.08]$, where $x_{TS}$ and $y_{TS}$ are the $x$- and $y$-coordinates of TS at a given time $t$, respectively. The energy conversion rates for the thermal and the kinetic energies are calculated in region $\Omega$ as below:
\begin{equation}
     \gamma_{heat} =\frac { E_{TL}-E_{TI}- E_{TF}}{ \Delta t\cdot m } ,
     \label{eq:et rate}
\end{equation}
and
\begin{equation}
      \gamma_{kine} =\frac { E_{KL}-E_{KI}- E_{KF}}{ \Delta t\cdot m } ,
      \label{eq:ek rate}
\end{equation}
where $E_{TL}$ and $E_{KL}$ are the thermal and the kinetic energies in $\Omega$ at time $t$, while $E_{TI}$ and $E_{KI}$ are the thermal and the kinetic energies confined in $\Omega$ at time $t-\Delta t$; $E_{TF}$ and $E_{KF}$ are the thermal and the kinetic energies flowing into $\Omega$; and $m$ is the total mass in $\Omega$ and $\Delta t$ is the time step for data sampling with $\Delta t=0.1$. More details about the computing approach can be found in \citet{Ni2012} and \citet{Ye2021}. Our results are given in Fig. \ref{fig:transferrate} for the time interval between 20 and 80.

Fig. \ref{fig:transferrate} indicates that before the flare loop and plasmoids appear, both rates remain quite close to 0. When the CS gets thinner and thinner at about $t=28$, the tearing mode instability occurs in the CS and accelerates the energy conversion. Downward outflows collide with the closed flare loop, producing TS at the top of flare loops \citep{Shen2018}. Once TS forms, both rates experience a jump and apparent energy accumulation starts. Similar processes of collisions between plasmoids and flare loops continued during the whole process and causes the successive increase in both kinetic and thermal energies. To compare the detailed accumulation of the thermal and kinetic energies from $t=20$ to $t=80$, we integrate the rate shown in Fig.\ref{fig:transferrate}\subref{fig:ene1} over this time interval. The results are plot in Fig. \ref{fig:transferrate}\subref{fig:ene2}. We notice that before the tearing mode instability takes place, the energy accumulation is at a low level. After $t=28$, both kinetic and thermal energies experience significant increase in accumulation. At $t=45$, the reconnection starts the fast phase and the accumulation rates reach a plateau. 

Fig. \ref{fig:transferrate}\subref{fig:ene2} also shows that the accumulative rates for thermal and kinetic energies possess the same trend. However, the rate of the increase in the thermal energy is about $4-5$ times of that in the kinetic one. The fact here that the thermal energy accumulates behind the TS more rapidly than the kinetic energy is consistent with the results of \citet{Murphy2011} and \citet{Ye2021}.

\section{Summary and Conclusions}
\label{sec:summary}

The 3D phenomenon occurring in the two-ribbon flare was investigated via 2D simulations in this work. This could be done because of the special geometric structure of the magnetic configuration involved in the solar eruption that produces the two-ribbon flare. As \citet{Lin2000} pointed out that the solar eruption is associated with thrusting of the flux rope, which apparently decreases the pressure in the region where the flux rope used to stay, and severely stretching of the magnetic field behind the flux rope, which develops a long CS through the low pressure region (refer to Figure 1 of \citet{Lin2005}). The difference in the pressure pushes both the magnetic field and the plasma toward the CS, which invokes the driven reconnection process in the CS that is obviously different from the spontaneous reconnection studied by \citet{Kowal2017,Kowal2020} and \citet{Beresnyak2017}. Furthermore, squeezing of the CS by the reconnection inflow confines all the processes occurring in the sheet to a very limited space in which the freedom in one direction is significantly suppressed. This implies that behaviors of any activities in such a sheet is inhomogeneous. Here, the inhomogeneity is not because of the existence of magnetic field, but due to the confinement by the reconnection inflow.

In this work, we focus on turbulent properties of the magnetic reconnection process in the CS and around the TS above the flare loop system. The Lundquist number of the system is $10^6$, and the grid resolution for calculation is high compared to those used used in previous works (e.g., see \citealt{shen2011,shen2013,Ye2021}).  Initially, magnetic reconnection commences in a large-scale Sweet-Parker CS. As reconnection progresses, the CS gradually gets thinner and thinner until the tearing mode instability is triggered. Plasmoids are formed inside the CS, bringing the reconnection process into the nonlinear phase. Turbulence leads to the fragmentation of the CS, the reconnection process manifests cascading behavior. Consequently, the fast mode of magnetic reconnection is switched on, and complex multi-scale features appear in the CS and the region between the CS and the flare loop. We carefully studied these features and looked into their physical properties. The main results are as follow:
\begin{enumerate}[(1)]
\item Magnetic reconnection continues to send plasma into the plasmoid. As getting heavy enough, an upward moving plasmoid above the PX-point may turn to fall down eventually, forcing both the PX-point and plasmoids below it to move downward together, and to merge into the flare loop system. The original PX-point structure is thus destroyed, an ordinary X-point above the heavy plasmoid upgrades to the PX-point almost instantaneously, and the CS configuration including the PX-point is renewed. This phenomenon and the associated process never occurs for the case without the gravity.

\item Following practice of previous works, we use term ``extra dissipation'' to describe any effective diffusion of magnetic field in numerical experiments in addition to the Spitzer resistivity. The contribution of the numerical diffusion to the extra dissipation remains unchanged once the algorithm and the code for calculations are given. The level of the extra dissiaption stays low before the tearing mode. Invoking of the tearing mode enhances the extra dissipation significantly within a short time. This explains why fast reconnection could still take place in a large-scale CME/flare CS.

\item The Taylor micro scale of the turbulence inside the CS, $l_{T}$, was found coincident with the CS thickness, $d$, which implies that the thickness of the CME/flare CS is governed by the Taylor micro scale.

\item Upward moving plasmoids are bigger than those moving downward because of the lower pressure at higher altitudes. Variations of the plasmoid number versus width and area manifest power law feature, $f\left( \psi  \right) \sim{  \psi }^{ \gamma }$, with indices, $\gamma$, of $-0.77$ and $-1.46$, respectively.

\item Three types of TS were recognized, including horizontal, V-like and trapezoid-like styles, in the cusp region above flare loop system. The turbulence could be strengthened by the TS. The more irregular and asymmetric the TS structure is, the stronger the enhancement is. The efficiency of energy transfer around the TS indicates that plasma heating is 5 times more efficient than accelerating, which is consistent with the result of previous works by \citet{Murphy2011} and \citet{Ye2021}.

\item Last but not the least, recent work in 3D by \citet{Jiang2021} on the solar eruption indicated that the reconnection process was accelerated apparently as the plasmoid instability occurs, and turbulent features in the reconnection region were found similar to what has been shown in the present work. In the future, we shall perform full 3D experiments for reconnection in the two-ribbon flare current sheet.
\end{enumerate}

\begin{acknowledgements}
We are grateful very much for the referee's valuable comments and suggestions that helped improve this article greatly. This work was supported by the Strategic Priority Research
Programme of Chinese Academy of Sciences (CAS) with
grants XDA17040507, and QYZDJ-SSWSLH012, the NSFC
grants 12073073, 11933009, 11973083, and U2031141, grants
associated with the Yunling Scholar Project of the Yunnan
Province and the Yunnan Province Scientist Workshop of
Solar Physics, and grants 202101AT070018 and 2019FB005 associated with the Applied Basic Research of Yunnan Province. Calculations in this work were performed on the cluster in the Computational Solar Physics Laboratory of the Yunnan Observatories.
\end{acknowledgements}

%\begin{thebibliography}{99}
%% you can type \apj for ApJ, \aap for A&A, \apss for Ap&SS, etc. Please consult
%% the macro chjaa.cls. You can also find them in aasguide.tex (AASTeX for ApJ, AJ, PASP)
%% Please follow the format of ChJAA's reference list

%\end{thebibliography}
\bibliographystyle{raa}
\bibliography{bibtex} 

\label{lastpage}

\end{document}